\documentclass{article} % For LaTeX2e
\usepackage{arxive_style,times}

\usepackage{amsmath,amsfonts,bm}

\def\eqref#1{equation~\ref{#1}}

\def\1{\bm{1}}

\DeclareMathAlphabet{\mathsfit}{\encodingdefault}{\sfdefault}{m}{sl}
\SetMathAlphabet{\mathsfit}{bold}{\encodingdefault}{\sfdefault}{bx}{n}

\usepackage{hyperref}
\usepackage{url}
\usepackage{listings}
\usepackage{xcolor}
\usepackage{adjustbox}
\usepackage{cleveref}
\usepackage{tikz}
\usepackage{pgfplots}
\pgfplotsset{compat=1.18}
\usepackage{booktabs}
\usepackage{tcolorbox}
\tcbuselibrary{listings,breakable,skins}
\usepackage{needspace}
\usetikzlibrary{automata,angles,quotes,calc, matrix, positioning,decorations.pathreplacing,shadows,shapes, arrows.meta}

\usepackage{enumitem}
\usepackage{subcaption}
\usepackage{graphicx}
\usepackage{wrapfig}
\usetikzlibrary{shapes.geometric}

\graphicspath{{debug_cwm_figures/}}

\definecolor{ipythonprompt}{RGB}{40,40,40}
\definecolor{ipythonout}{RGB}{200,0,0}
\definecolor{pythonstring}{RGB}{186,85,211}
\definecolor{pythonkeyword}{RGB}{0,0,255}
\definecolor{pythoncomment}{RGB}{128,128,128}

\definecolor{modernblue}{RGB}{52, 152, 219}
\definecolor{darkgreen}{RGB}{46, 204, 113}
\definecolor{modernorange}{RGB}{230, 126, 34}
\definecolor{modernpurple}{RGB}{155, 89, 182}
\definecolor{modernred}{RGB}{231, 76, 60}
\definecolor{modernyellow}{RGB}{241, 196, 15}
\definecolor{moderngray}{RGB}{149, 165, 166}
\definecolor{modernlightgray}{RGB}{236, 240, 241}
\definecolor{moderndarkgray}{RGB}{52, 73, 94}
\definecolor{navyblue}{RGB}{28, 57, 93}
\definecolor{darkgreen}{RGB}{27, 94, 32}  % deep green, print-friendly

\newtcblisting{cwmtrace}{
  enhanced,
  breakable,
  listing only,
  listing engine=listings,
  colback=white,
  colframe=white,
  boxrule=0pt,
  frame hidden,
  sharp corners,
  borderline west={0.9pt}{0pt}{navyblue}, % thinner left bar
  left=6pt, right=0pt, top=2pt, bottom=2pt,
  boxsep=0pt,
  before skip=4pt,
  after skip=4pt,
  listing options={
    basicstyle=\ttfamily\scriptsize, % ↓ from \footnotesize
    breaklines=true,
    breakatwhitespace=false,
    columns=fullflexible,
    keepspaces=true,
    showstringspaces=false,
    frame=none
  }
}

\title{Debugging Code World Models}

\author{
Babak Rahmani\thanks{This work was conducted while at Microsoft Research.}\\
T\"ubingen AI Center, University of T\"ubingen, 
Microsoft Research\\
\texttt{rahmani.b91@gmail.com}
}

\iclrfinalcopy % Uncomment for camera-ready version, but NOT for submission.
\begin{document}

\maketitle

% \vspace{-3em}
% \begin{center}
% \small
% \href{https://babak70.github.io/code-world-models-blog/assets/reports/cruxeval_failure_report.html}{ \textbf{CruxEval}} \quad
% \href{https://babak70.github.io/code-world-models-blog/assets/reports/humaneval_failure_report.html}{ \textbf{HumanEval}} \quad
% \href{https://babak70.github.io/code-world-models-blog/assets/reports/nesting_report.html}{ \textbf{Composition}}
% \end{center}
% \vspace{1em}

\vspace{-3em}
\begin{center}
\small
\href{https://babak70.github.io/code-world-models-blog}{\textbf{[Blog post]}} \quad $\vert$ \quad
\href{https://babak70.github.io/code-world-models-blog/assets/reports/cruxeval_failure_report.html}{\textbf{[CruxEval report]}} \quad
\href{https://babak70.github.io/code-world-models-blog/assets/reports/humaneval_failure_report.html}{\textbf{[HumanEval report]}} \quad
\href{https://babak70.github.io/code-world-models-blog/assets/reports/nesting_report.html}{\textbf{[Composition report]}}
\end{center}
\vspace{1em}

\begin{abstract}
Code World Models (CWMs) are language models trained to simulate program execution by predicting explicit runtime state after every executed command. This execution-based world modeling enables internal verification within the model, offering an alternative to natural language chain-of-thought reasoning. However, the sources of errors and the nature of CWMs' limitations remain poorly understood. We study CWMs from two complementary perspectives: local semantic execution and long-horizon state tracking. On real-code benchmarks, we identify two dominant failure regimes. First, dense runtime state reveals produce token-intensive execution traces, leading to token-budget exhaustion on programs with long execution histories. Second, failures disproportionately concentrate in string-valued state, which we attribute to limitations of subword tokenization rather than program structure. To study long-horizon behavior, we use a controlled permutation-tracking benchmark that isolates state propagation under action execution. We show that long-horizon degradation is driven primarily by incorrect action generation: when actions are replaced with ground-truth commands, a Transformer-based CWM propagates state accurately over long horizons, despite known limitations of Transformers in long-horizon state tracking. These findings suggest directions for more efficient supervision and state representations in CWMs that are better aligned with program execution and data types.
\end{abstract}

\section{Introduction}
%==============================================================================

\begin{wrapfigure}{r}{0.35\columnwidth}
\vspace{-20pt}
\centering
\includegraphics[width=\linewidth, trim=7 10 10 0, clip]{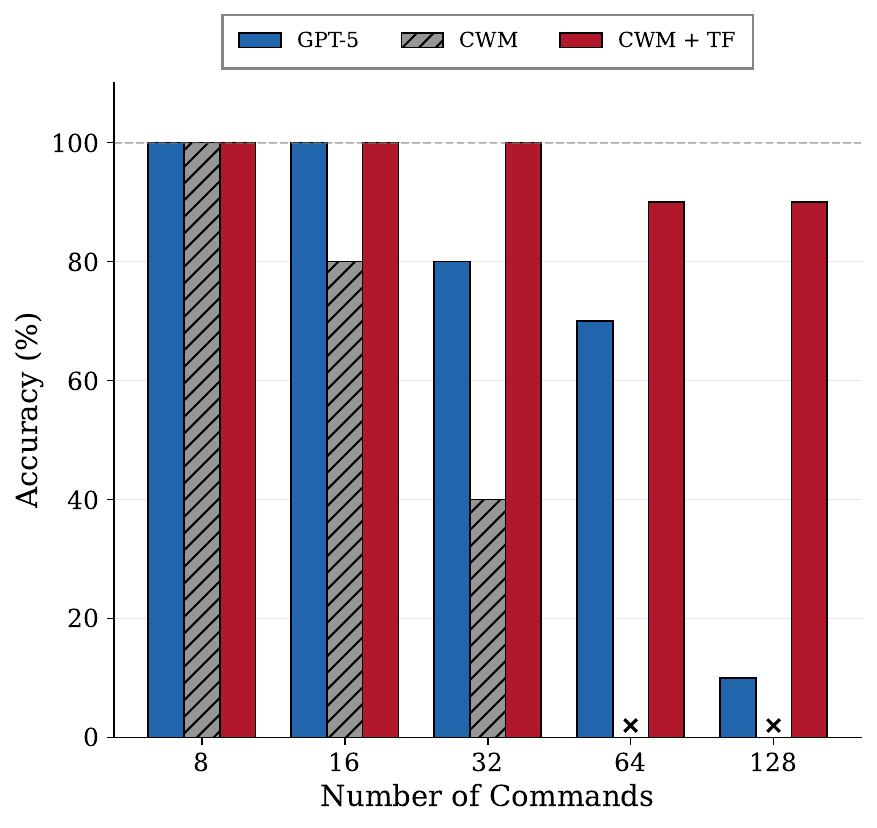}

% \caption{Accuracy on long-horizon state tracking via Code $S_5$ permutation tracking across sequence lengths (8--128 swaps). 
% CWM+TF (teacher forcing) maintains high accuracy while baselines GPT5 and CWM degrade. 
% $\times$ indicates zero accuracy.}
\caption{Accuracy on long-horizon state tracking via Code $S_5$ permutation tracking, where models apply sequences of permutation swaps (8--128 operations). 
CWM+TF (teacher forcing) maintains high accuracy, while GPT5 and CWM degrade with sequence length. 
$\times$ indicates zero accuracy.}
\vspace{-30pt}
\label{fig:main_results}
\vspace{5pt}
\end{wrapfigure}

World modeling frameworks recast sequence modeling as learning an explicit state-transition process: given a current state and an action, the model predicts how the environment evolves~\citep{ha2018world,bruce2024genie,lehrach2025code,dainese2024generating,tang2024worldcoder,kim2022plm}. 
This formulation transforms LLMs from passive predictors into simulators of structured dynamics which could help with supporting multi-step reasoning, planning, and internal verification through rollouts.

Recent work extends this world-modeling paradigm to code~\citep{liu2023code,ding2024semcoder,ding2024traced,lehrach2025code}, where the environment is program execution and the state consists of runtime variables and control-flow state. 
Code World Models (CWMs)~\citep{copet2025cwm} instantiate this approach by training on Python execution traces that interleave each executed line of code with a complete snapshot of the runtime state. 
This yields a dense supervision regime in which all variables are revealed after every operation. 
CWMs have been shown to deliver substantial gains on software engineering benchmarks~\citep{copet2025cwm}.

Despite these gains, it remains unclear why dense action–state prediction is so effective and where its limitations arise. Revealing the full runtime state after every operation reduces execution to a sequence of locally supervised state updates, effectively mitigating the known long-horizon state-tracking difficulties of Transformer-based models. But how fully does this supervision resolve long-horizon failures in practice? And even under dense reveals, can errors still arise from local semantic execution?

In this work, we study CWMs through these two complementary lenses. First, we evaluate semantic execution, asking whether models correctly apply sequences of deterministic operations across data types. Second, we analyze long-horizon state tracking, asking whether execution state can be maintained over many steps despite compounding autoregressive error and limited context.

\textbf{Contributions}. We begin with real-code benchmarks, CruxEval-O~\citep{gu2024cruxeval} and HumanEval~\citep{chen2021evaluating}, where we observe failures concentrate in two regimes (\cref{sec:bench_failure_analysis}): \textbf{(1)} token-budget exhaustion from long execution traces and \textbf{(2)} brittleness in string-valued state. The latter aligns with known representation discontinuities induced by subword tokenization~\citep{xue2021byt5,clark2022canine,pagnoni2025blt}. 

To isolate data type effects independent of program structure, we introduce controlled evaluations. Using functional programs with fixed structure and varying data types, we show that CWMs compose reliably for non-string data, while nested string transformations degrade sharply (\cref{sec:compositionality,sec:tokenization}).

To isolate long-horizon behavior, we study an existing controlled permutation-tracking benchmark~\citep{siems2026learning} that stresses state propagation over extended execution histories (\cref{sec:state_tracking}). Despite full state reveals at every step, performance degrades with horizon length. We show that this degradation is driven primarily by action hallucination rather than local state-update errors: when actions are corrected, a Transformer-based CWM propagates state accurately for over 128 steps.

% Finally, motivated by efficiency and truncation concerns, we vary reveal spacing to study reduced state supervision (\Cref{sec:state_tracking}). 
% Transformer-based CWMs degrade rapidly as supervision becomes sparse, whereas recurrent architectures remain stable, indicating that dense supervision compensates for architectural limitations of Transformer-based CWMs.
% We open-source our analysis of failure modes across four datasets.

Finally, motivated by efficiency and truncation concerns, we discuss the implications of dense state supervision for scalability in code world modeling (\cref{sec:discussion}). While we do not resolve these trade-offs here, our analysis highlights the need for future work that studies execution under reduced observation. We will open-source our analysis across all datasets in the future.

\section{Background}
\vspace{-5pt}
\textbf{Code World Models and execution traces}.
A world model learns a transition function for a sequential environment: given a state and an action, it predicts the next state~\citep{ha2018world}.
In code execution, actions correspond to executed statements and the state corresponds to the runtime configuration, including variable bindings and control-flow state.

CWMs instantiate this by representing program execution as an explicit sequence of action--state pairs derived from Python execution traces~\citep{copet2025cwm}.
Given a partial trace, the model is trained to predict the next action and the resulting state, effectively acting as a simulator of program execution. For example:

% \paragraph{Example: execution trace.}
% Consider the following program:

% \begin{cwmtrace}
% def f(x):
%     y = x + 1
%     return y

% def main():  # << START_OF_TRACE
%     z = f(3)
%     return z
% \end{cwmtrace}

% \begin{cwmtrace}
% <|frame_sep|><|call_sep|>{}<|action_sep|>def main():  # << START_OF_TRACE
% <|frame_sep|><|line_sep|>{}<|action_sep|>    z = f(3)
% <|frame_sep|><|call_sep|>{"x": "3"}<|action_sep|>def f(x):
% <|frame_sep|><|line_sep|>{"x": ".."}<|action_sep|>    y = x + 1
% <|frame_sep|><|line_sep|>{"x": "..", "y": "4"}<|action_sep|>    return y
% <|frame_sep|><|return_sep|><|action_sep|>    return y
% <|arg_sep|>"4"
% <|frame_sep|><|line_sep|>{"z": "4"}<|action_sep|>    return z
% <|frame_sep|><|return_sep|><|action_sep|>    return z
% <|arg_sep|>"4"
% <|frame_sep|>

% \end{cwmtrace}

\begin{cwmtrace}
# Program
def f(x):
    y = x + 1
    return y

def main():  # << START_OF_TRACE
    z = f(3)
    return z

# Execution trace (action -> state)
Action: def main()                State: {}
Action: z = f(3)                  State: {}
Action: def f(x)                  State: {"x": 3}
Action: y = x + 1                 State: {"x": .., "y": 4}
Action: return y                  State: {"x": .., "y": ..}   -> returns 4
Action: return z                  State: {"z": 4}           -> returns 4
\end{cwmtrace}

At each step, the model observes the executed statement (action) and a complete snapshot of the runtime state, where "$..$" denotes variables that remain unchanged. CWMs are trained to predict the next action–state pair given a code prefix.
Full trace examples are provided in Appendix~\ref{app:humaneval_cruxeval_prompts}.

% \paragraph{State tracking as a long-horizon diagnostic}
% State tracking refers to the ability of a model to maintain and update an internal representation of system state across a sequence of operations~\citep{liu2022transformers}.
% Rather than viewing code execution as a single input--output mapping, we analyze CWMs through this lens, asking whether failures arise from incorrect state propagation, incorrect action generation, unstable representations, or limitations imposed by the supervision structure.
% This perspective motivates the controlled benchmarks and analyses in ~\cref{sec:state_tracking} .

\section{Evaluation on Code Benchmarks}

We evaluate CWMs on real code execution benchmarks to identify recurring failure modes and their underlying causes, and then use controlled experiments (\cref{sec:compositionality,}) to isolate these effects. 

We begin with two standard benchmarks that probe complementary aspects of code execution. CruxEval-O~\citep{gu2024cruxeval} evaluates end-to-end output prediction for short Python programs, emphasizing direct reasoning from source code to final outputs. HumanEval~\citep{chen2021evaluating} assesses functional correctness via unit tests. In our setting, both benchmarks are treated as execution probes: given a fixed program and input, the model must simulate execution to produce the correct output. Example prompt templates are provided in Appendix~\ref{app:humaneval_cruxeval_prompts}. Under this setup, CWM achieves 85.0\% accuracy on CruxEval-O and 91.4\% on HumanEval.

\begin{table}[t]
\centering
\caption{CWM accuracy on code execution benchmarks. \textcolor{darkgreen}{$\Delta$} denotes absolute accuracy change after intervention.}

\label{tab:benchmark_results}
\footnotesize
\setlength{\tabcolsep}{3pt}
\begin{tabular}{lccccccc}
\toprule
\textbf{Benchmark} & \textbf{Samples} & \textbf{Max Tokens} & \textbf{Correct} & \textbf{Incorrect} & \textbf{Truncated} & \textbf{After Intervention} & \textbf{\textcolor{darkgreen}{$\Delta$}} \\
\midrule
CruxEval-O & 800 & 8K & 85.0\% & 14.2\% & 0.8\% & \textbf{90.4\%} & \textcolor{darkgreen}{+5.4} \\
HumanEval & 723 & 8K & 91.4\% & 4.5\% & 4.1\% & \textbf{92\%} & \textcolor{darkgreen}{+0.6} \\
\midrule
\multicolumn{8}{l}{\textit{Composition Zoo (depth=5, non-string)}} \\
\quad Boolean & 10 & 8K & 100\% & 0\% & --- & --- & --- \\
\quad Bitwise & 10 & 8K & 100\% & 0\% & --- & --- & --- \\
\quad Math & 10 & 8K & 100\% & 0\% & --- & --- & --- \\
\quad Character & 10 & 8K & 100\% & 0\% & --- & --- & --- \\
\quad List & 10 & 8K & 100\% & 0\% & --- & --- & --- \\
\quad Set & 10 & 8K & 100\% & 0\% & --- & --- & --- \\
\quad Dictionary & 10 & 8K & 100\% & 0\% & --- & --- & --- \\
\midrule
\multicolumn{8}{l}{\textit{Composition (string)}} \\
\quad depth=2 & 100 & 4K & 75\% & 25\% & --- & \textbf{78\%} & \textcolor{darkgreen}{+3} \\
\quad depth=3 & 100 & 4K & 58\% & 42\% & --- & \textbf{63\%} & \textcolor{darkgreen}{+5} \\
\quad depth=4 & 100 & 4K & 39\% & 61\% & --- & \textbf{43\%} & \textcolor{darkgreen}{+4} \\
\quad depth=5 & 100 & 4K & 25\% & 75\% & --- & \textbf{28\%} & \textcolor{darkgreen}{+3} \\
\bottomrule
\end{tabular}
\end{table}

\Cref{fig:failure_combined}, top shows the distribution of non-truncation failures (i.e., incorrect predictions) by output data type. 
String-valued outputs dominate failure cases in both benchmarks. 
In CruxEval-O, strings account for 73\% of failures despite constituting only 46\% of outputs (1.58$\times$ overrepresentation), while integers (11\%), lists (9\%), and other types fail far less frequently. 
HumanEval shows a similar pattern: string-related failures comprise 44\% of failure cases even though strings represent just 17\% of outputs (2.58$\times$ overrepresentation). 
Full type distributions are provided in Appendix~\ref{app:datatype_distribution}.

To contextualize these errors, \cref{fig:failure_combined}, bottom presents a taxonomy of failure modes on CruxEval-O, separating semantic prediction errors from trace truncation due to token-budget exhaustion. We next focus on these two failure categories.

\subsection{Where Code World Models Fail?}
\label{sec:bench_failure_analysis}

\begin{figure}[t]
\centering
\includegraphics[width=\textwidth]{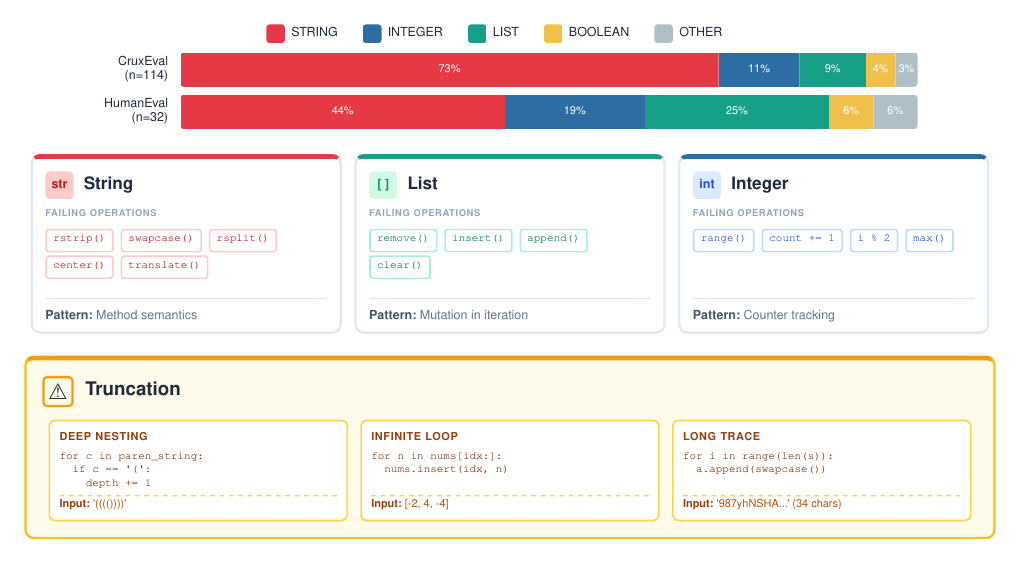}
\caption{\textbf{Top:} Distribution of CWM non-truncation failures by output data type on CruxEval-O and HumanEval, excluding truncation cases. String-valued outputs dominate failures \textbf{Bottom:} Failure taxonomy on CruxEval-O. Top row shows examples of data type failures. Bottom row shows truncation failure patterns that cause trace overflow.}
\label{fig:failure_combined}
\end{figure}

\Cref{fig:failure_combined}, bottom summarizes failure modes on CruxEval-O, grouped by output data type and truncation. A complete catalog of failure cases, including code, inputs, and predictions, is provided in Appendix~\ref{app:failure_catalog}. We focus on two dominant failure families.

\paragraph{Failure family 1: trace truncation.}
Although both benchmarks consist largely of short functions, dense trace supervision can fail when execution induces traces that exceed the token budget.
As illustrated in \cref{fig:failure_combined}, truncation commonly arises from (i) deep but linear iteration chains (e.g., repeated parenthesis depth tracking), (ii) control-flow patterns that generate unbounded execution through in-place mutation during iteration, and (iii) per-element string processing that produces long, token-expensive traces even for moderate input lengths.
In these cases, failure is not caused by incorrect local execution but by the cumulative cost of revealing state at every step.

% A related phenomenon is \emph{loop drift}, where execution remains within the context window but the model becomes inconsistent in maintaining loop-carried state across many iterations.
% Both truncation and drift stem from long-horizon dependencies induced by dense state reveals under iterative control flow, rather than from semantic ambiguity in individual operations.

\paragraph{Failure family 2: string-valued state.}
Beyond truncation, failures disproportionately involve string-valued state.
As illustrated in \cref{fig:failure_combined}, these errors are dominated by incorrect handling of common string operations, including method semantics (e.g., \texttt{rstrip}, \texttt{swapcase}, \texttt{rsplit}, \texttt{center}) and boundary-sensitive indexing or slicing.
The resulting outputs are often syntactically well-formed but semantically incorrect, indicating failures in how string transformations are represented rather than in control flow.

Notably, these errors arise even in short programs with minimal execution depth, and similar string-related failures appear in HumanEval.
This raises a central question: do these failures reflect limitations in CWMs' semantic execution, or instability in how string state is represented under tokenization?

\begin{tcolorbox}[
    colback=modernlightgray,
    colframe=navyblue,
    boxrule=0.6pt,
    arc=2pt,
    left=6pt,
    right=6pt,
    top=6pt,
    bottom=6pt
]
\textbf{Takeaway.}
On real code benchmarks, CWM failures under dense supervision concentrate in two regimes: \emph{(i)} token-budget limitations from long execution traces, and \emph{(ii)} representation brittleness in string-valued state.
\end{tcolorbox}

\subsection{Functional composition across data types}
\label{sec:compositionality}

To isolate the source of string-related failures observed in real-code benchmarks, we introduce a controlled test based on functional composition. Composing deterministic single-argument functions to depth $d$ induces multi-step computation without loops or branching, allowing us to hold program structure fixed while varying only the data type.

\paragraph{Composition Zoo (non-string data types).}
We first evaluate a multi-domain Composition Zoo consisting of depth-5 compositions across seven non-string categories (boolean, bitwise, math, character, list, set, dictionary), with 10 held-out samples per category. Functions are deterministic and side-effect free, and compositions are generated by randomly nesting functions within each category. At depth 5, CWM achieves 100\% accuracy across all categories, demonstrating reliable composition when intermediate state is represented stably. Appendix~\ref{app:composition_zoo_function_defs} lists the function definitions.

\paragraph{Composition (string).}
We next apply the same compositional scaffold to string transformations using the composition string dataset from~\citep{yuan2025composition}. The dataset contains deterministic string-manipulation functions, including case alternation, prefix/suffix insertion, joining, slicing, rotation, and simple loop-based transformations. The full function set is listed in Appendix~\ref{app:composition_zoo_function_defs}.

In contrast to the non-string results, accuracy degrades sharply with composition depth (Table~\ref{tab:benchmark_results}), despite programs being short, deterministic, and purely functional. To factor out per-operation difficulty, we first evaluate atomic calls (depth 1) for 25 string functions (10 samples each) and retain the 15 functions with $\geq$90\% atomic accuracy, yielding an average atomic accuracy of 95.3\%. Per-function accuracy is reported in Appendix~\ref{app:atomic_string_functions}.
We then generate 100 random compositions per depth for $d \in \{1,2,3,4,5\}$, with ground-truth outputs computed by executing the Python code. A representative evaluation prompt is shown in Appendix~\ref{app:comp_prompt}. Performance drops from 75\% at depth 2 to 25\% at depth 5, in stark contrast to non-string compositions, which maintain 100\% accuracy at depth 5 under identical structure.

These results support the hypothesis that the primary limitation lies in string representation rather than execution structure, motivating the analysis in~\cref{sec:tokenization}.

\begin{tcolorbox}[
    colback=modernlightgray,
    colframe=navyblue,
    boxrule=0.6pt,
    arc=2pt,
    left=6pt,
    right=6pt,
    top=6pt,
    bottom=6pt
]
\textbf{Takeaway.}
Functional composition is not a bottleneck for CWMs. When program structure is held fixed, CWMs compose reliably across non-string data types but degrade sharply on string-valued state, isolating string representation as the source of failure.
\end{tcolorbox}

\subsection{Semantics-Preserving Code Decomposition Interventions}
\label{sec:intervention}
To probe even further whether some CWM failures arise from hidden intermediate state rather than incorrect execution, we apply a simple code decomposition intervention to failing samples across all dataset.
The intervention rewrites programs using semantics-preserving transformations that expose intermediate values explicitly in the execution trace, including (i) decomposition of nested expressions into temporary assignments and (ii) decomposition of selected single-character string methods into explicit character-level loops.
This yields consistent but limited improvements on the observed failure cases (Table~\ref{tab:benchmark_results}, ``After Intervention''), indicating that a subset of errors is driven by state visibility.
However, the approach is not scalable: string decompositions can inflate trace length and trigger truncation, and decomposition does not address failures caused by semantic errors.
Implementation details and examples are deferred to the Appendix~\ref{app:intervention}.

\subsection{Tokenization discontinuity}
\label{sec:tokenization}

The composition experiments show that string-valued state is uniquely brittle under depth, even when program structure is held fixed. This behavior is consistent with tokenization discontinuity, where the same character sequence produces different tokens depending on surrounding context.

\begin{wrapfigure}{r}{0.6\columnwidth}
\vspace{-15pt}
\centering
\includegraphics[width=\linewidth, trim=10 300 25 0, clip]{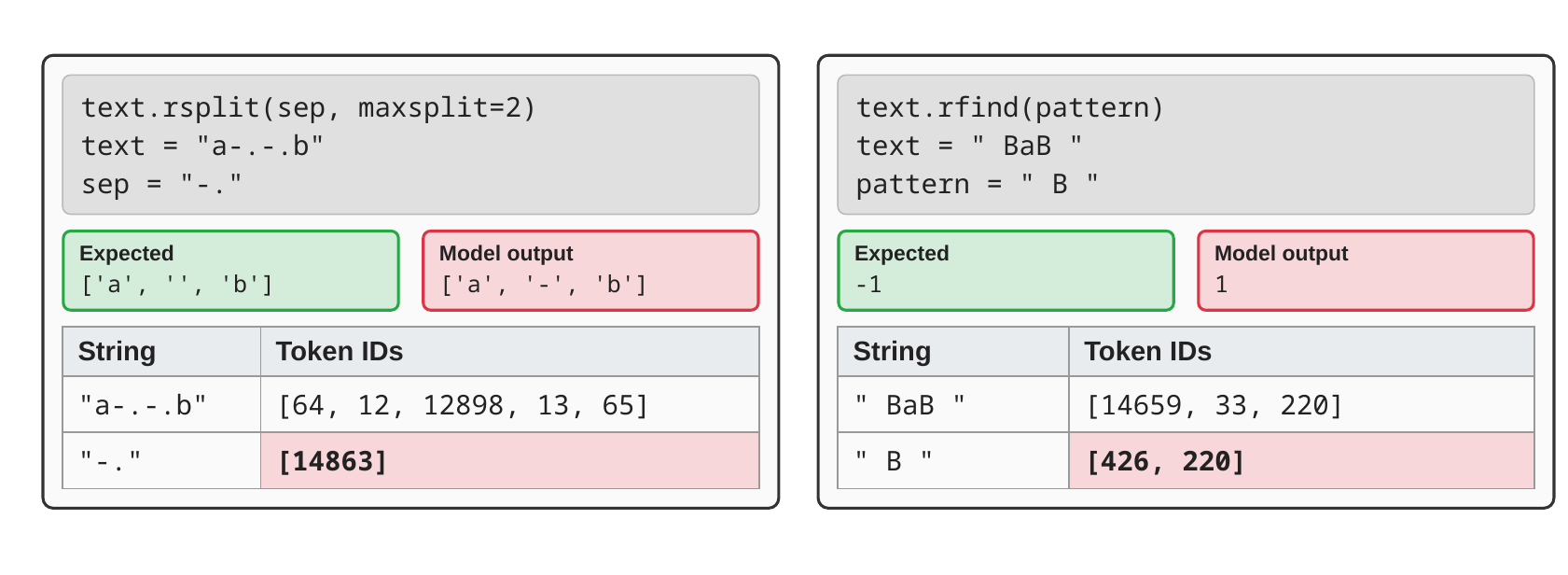}
\vspace{-15pt}
\caption{Tokenization discontinuity: \textbf{Left}: the separator \texttt{"-."} tokenizes as ID 14863 alone, but this token never appears in \texttt{"a-.-.b"}'s token sequence, causing \texttt{rsplit} to fail. \textbf{Right}: the pattern \texttt{" B "} tokenizes as \texttt{[426, 220]}, but token 426 is absent from \texttt{" BaB "}'s tokens, causing \texttt{rfind} to hallucinate a match.}
\label{fig:tokenization_discontinuity}
\vspace{-10pt}
\end{wrapfigure}

Under Byte-Pair Encoding (BPE)~\cite{sennrich2016neural} tokenization, context-dependent merging causes the same characters to map to different token IDs. As shown in \cref{fig:tokenization_discontinuity}, the separator \texttt{"-."} tokenizes as a single token (ID 14863) when encoded alone, but within \texttt{"a-.-.b"}, BPE produces an entirely different sequence: \texttt{[64, 12, 12898, 13, 65]}. Critically, token 14863 never appears in this sequence. Consequently, when evaluating \texttt{rsplit}, the model cannot locate the separator at the token level, producing incorrect output.

Similarly, for \texttt{rfind(" B ", " BaB ")}, the pattern tokenizes as \texttt{[426, 220]}, but the text yields \texttt{[14659, 33, 220]} where token 426 is absent. The model hallucinates a match at position 1 when the correct answer is $-1$. 

This instability explains the sharp error compounding observed in string compositions. Each transformation can substantially perturb the token sequence, making subsequent state updates unreliable even in short, deterministic programs. 
% In contrast, non-string data types admit stable representations, consistent with their perfect compositional performance.

\section{Long-Horizon State Tracking: The Code \texorpdfstring{$S_5$}{S5} Benchmark}
\label{sec:state_tracking}

Beyond representation-specific failures such as string brittleness, a central question is whether CWMs can faithfully track execution state over long horizons. This question is orthogonal to local semantic execution and arises even when code semantics are simple, deterministic, and fully observed, as in CWMs with full state reveals at every step.

CWMs are implemented using attention-based architectures. Prior work shows that Transformer-style models struggle with faithful state tracking and automaton simulation as sequence length grows, even when trained successfully on short horizons~\citep{hahn2020theoretical,bhattamishra2020ability,deletang2022neural,merrill2023parallelism}. 
% To evaluate this limitation in isolation, we introduce a controlled state-tracking benchmark based on permutation groups expressed as executable code~\citep{siems2026learning}.

To evaluate this limitation in isolation, we adopt a controlled state-tracking benchmark based on permutation groups expressed as executable code, introduced by \citet{siems2026learning}.

\paragraph{Permutation tracking on $S_n$.}
Let $S_n$ denote the symmetric group on $n$ elements. Given a sequence of permutations $(\sigma_1, \dots, \sigma_N)$, the cumulative state after $t$ steps is
\[
\sigma_{\le t} = \sigma_t \circ \sigma_{t-1} \circ \cdots \circ \sigma_1 \in S_n.
\]
Equivalently, initializing from $x_0=(1,\dots,n)$, the tracked state is $x_t = \sigma_{\le t}(x_0),$ and the task is to predict the final state $x_N$ after $N$ operations.

To mirror code execution, we serialize permutations as Python variable assignments over $n$ variables, using full permutations that simultaneously reassign all $n$ variables. An example of $S_5$ expressed in code over five variables is shown in \cref{fig:teacher_forcing_process}, left.

\subsection{Code Execution State Tracking in Code World Models}

We evaluate long-horizon state tracking using permutation-tracking traces over $S_5$. Models must iteratively apply deterministic transitions and predict the final variable assignment after $N \in \{8,16,32,64,128\}$ steps. This benchmark isolates state propagation over long horizons. We report exact-match accuracy of the final state.

CWM is evaluated in its native execution-trace format, and GPT-5 is evaluated on the same traces as a general-purpose LLM baseline. Full prompt details are provided in Appendix~\ref{app:llm_evaluation_prompts}. Results are shown in \cref{fig:main_results}.

\begin{figure}[t] \centering \includegraphics[width=0.99\linewidth, trim=0 20 0 0, clip]{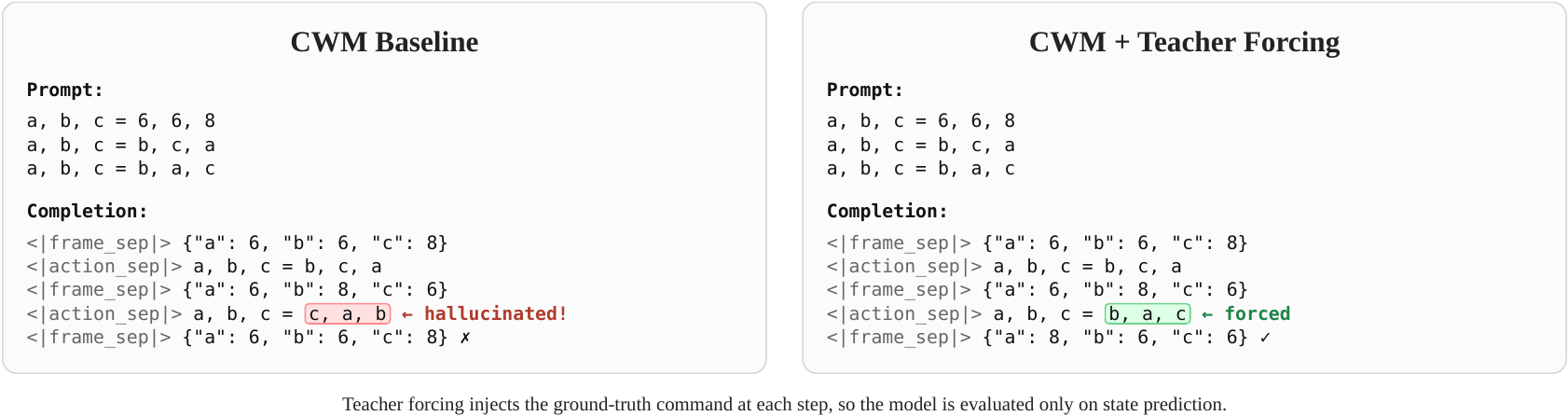} \caption{Illustration of action hallucination and teacher forcing in a CWM trace. \textbf{Left}: In the baseline setting, an incorrect next command corrupts the execution history and forces all subsequent states to be wrong. \textbf{Right}: Under teacher forcing, the ground-truth command is injected at each step and evaluation isolates state prediction.} \label{fig:teacher_forcing_process} \end{figure}

At first glance, CWM accuracy degrades sharply with horizon length, suggesting a failure of state propagation. However, inspection reveals a different dominant failure mode: errors typically arise from generating an incorrect next command (see \cref{fig:teacher_forcing_process}, left), after which all subsequent states are deterministically corrupted.

\subsection{Action Hallucination vs.\ State Propagation}

In the baseline setting, long-horizon failures are dominated by action hallucination: once an incorrect command is generated (e.g., an incorrect swap), all subsequent states are necessarily wrong because execution is conditioned on a corrupted history (see \cref{fig:teacher_forcing_process}).

To isolate state propagation from action generation, we perform a \textbf{teacher-forcing} evaluation (see \cref{fig:teacher_forcing_process}, right), injecting the ground-truth operation at each step and evaluating only the predicted state. Under teacher forcing, CWM maintains 90\% accuracy at 128 steps, whereas baseline accuracy collapses beyond 64 steps (\cref{fig:main_results}).

This demonstrates that, conditioned on correct actions, a Transformer-based CWM can reliably propagate state over long horizons. Crucially, this capability is enabled by dense full-state reveals, which reduce long-horizon tracking to a sequence of locally supervised state updates.

\begin{tcolorbox}[
    colback=modernlightgray,
    colframe=navyblue,
    boxrule=0.6pt,
    arc=2pt,
    left=6pt,
    right=6pt,
    top=6pt,
    bottom=6pt
]
\textbf{Takeaway.}
Long-horizon failures in CWMs are dominated by action hallucination rather than state-update errors. When conditioned on correct actions, Transformer-based models can faithfully propagate state for hundreds of steps. Dense full-state supervision is the key mechanism enabling this behavior.
\end{tcolorbox}

% \begin{figure}[t]
% \centering
% \begin{subfigure}[t]{0.60\textwidth}
%     \centering
%     \includegraphics[width=\linewidth]{repl_trace_style_swap_vs_fullassign.pdf}
% \end{subfigure}
% \hfill
% \begin{subfigure}[t]{0.38\textwidth}
%     \centering
%     \includegraphics[width=\linewidth]{figures/motivation/swaps.pdf}
% \end{subfigure}
% \caption{Left: state tracking as REPL traces (Swaps vs Full Permutation). Right: reveal spacing as a control knob: frequent full reveals reduce effective tracking horizon, while sparse reveals induce long-horizon tracking.}
% \label{fig:reveal_spacing_schematic}
% \end{figure}

%==============================================================================
\section{Related Work}
%==============================================================================

\textbf{Trace-supervised execution models.}
A growing line of work augments code models with execution signals to capture dynamic semantics.
CodeExecutor~\citep{liu2023code} uses mutation-based augmentation to generate large-scale programs with line-by-line variable traces.
SEMCODER~\citep{ding2024semcoder} trains on execution monologues that describe control-flow and state evolution in natural language.
TRACED~\citep{ding2024traced} combines source code with quantized execution states for improved value/path prediction.
At larger scale, CWMs~\citep{copet2025cwm} train on action--state traces where each executed line is paired with an explicit runtime snapshot, and report strong gains on software engineering benchmarks. We focus on CWMs and study their performance.

\textbf{When do traces help (and when do they not)?}
Recent studies question whether simply adding traces reliably improves general code reasoning.
\citet{zhang2025grounding} find traced versions of human-written functions can \emph{hurt} performance, suggesting models may pattern-match on trace format rather than internalize execution.
\citet{wang2025code} and \citet{haque2025towards} report mixed benefits from trace-augmented fine-tuning across model families, and highlight that traces can be more useful as a test-time tool than as a universal training signal.
\citet{armengol2025cannot} show that models can produce long scratchpads with high trace accuracy while yielding limited gains on downstream code generation.
These results motivate our focus on diagnosing trace-based execution: we characterize dominant failure modes on real benchmarks and isolate their causes with controlled tests.

\textbf{State tracking, supervision density, and architectural limits.}
Our analysis connects trace-based execution to classic state-tracking benchmarks studied through regular languages, automata emulation, and length generalization~\citep{hahn2020theoretical,bhattamishra2020ability,deletang2022neural,merrill2023parallelism}. 
A recurring finding is that Transformer-based models can fit short-horizon training distributions yet struggle to extrapolate faithful state updates to longer horizons~\citep{liu2022transformers}. 
Dense trace supervision alters this regime by revealing the full execution state after every operation, effectively reducing long-horizon state tracking to a sequence of locally supervised updates. 
At the same time, prior work shows that recurrent linear and state-space architectures can maintain stable state representations under sparse observation and extended horizons~\citep{orvieto2023resurrecting, grazziunlocking,siems2025deltaproduct,schoneimplicit}, highlighting an architectural dependence of state tracking on both supervision density and model class. 
Our results on code $s_5$ benchmark suggest that dense state supervision in CWMs plays a role analogous to the inductive bias provided by linear recurrent architectures, compensating for limitations in Transformer-based state propagation.

% We build on this lens by treating execution traces as \emph{state reveal supervision} and explicitly varying reveal spacing in a controlled permutation-tracking task, exposing an architectural dependence of Transformers on dense state reveals.

\section{Discussion and Conclusion}
\label{sec:discussion}

Our results highlight the limitations of CWMs. 
In the short-horizon regime, dense action--state supervision makes execution largely a sequence of locally supervised state updates, enabling strong compositional performance across most data types (\cref{sec:compositionality}). 
However failures concentrate in string-valued state, consistent with tokenization-driven representation brittleness (\cref{sec:tokenization,sec:bench_failure_analysis}). 
In the long-horizon regime, performance degradation is dominated by action hallucination: once an incorrect command is generated, the remainder of the trace is deterministically corrupted; when we correct actions via teacher forcing, CWMs can propagate state accurately (\cref{sec:state_tracking}). 
Across both regimes, the approach is inherently token-intensive, and long traces lead to truncation, making efficiency a central bottleneck (\cref{sec:bench_failure_analysis}).

The efficiency bottleneck, introduced by dense execution traces, also points to an architectural consideration. When state information is revealed less frequently, successful execution requires models to propagate state over extended horizons with limited observation. Recent concurrent work~\citep{siems2026learning}
 shows that linear recurrent architectures can maintain stable state tracking under sparse reveals, whereas Transformer-based models degrade more rapidly in this regime. This suggests that dense supervision may compensate for architectural limitations in attention-based simulators, and that reducing token cost fundamentally alters the learning regime for code execution.

Looking ahead, building world-modeling agents that execute and verify code internally, rather than relying on external tools, will require rethinking both supervision and architecture to improve efficiency.
Moving beyond dense execution traces calls for models that can propagate state under sparse observation, where linear recurrent~\citep{yanggateddeltanet,siems2025deltaproduct} or state-space architectures~\citep{gumamba} are natural candidates.
At the same time faithful execution will also require revisiting text representation itself, for example through byte-level or tokenizer-free approaches that avoid brittle subword boundaries and support stable character-level computation~\citep{choe2019bridging,xue2021byt5,clark2022canine,pagnoni2025blt}.

\bibliography{debug_cwm}
\bibliographystyle{iclr2025_conference}

\clearpage
\appendix

%===============================================================================
\section*{Appendix Overview}
%===============================================================================

\begin{itemize}[nosep,leftmargin=*]
    \item \textbf{Appendix~\ref{app:humaneval_cruxeval_prompts}: HumanEval and CruxEval: Prompt Examples}
    \item \textbf{Appendix~\ref{app:atomic_accuracy}: Compositionality: Additional Details}
    \item \textbf{Appendix~\ref{app:llm_evaluation_prompts}: Code $S_5$ Long-Horizon State-Tracking: Experiment Setup and Prompts (GPT-5 and CWM)}
    \item \textbf{Appendix~\ref{app:datatype_distribution}: Data Type Distribution Analysis}
    \item \textbf{Appendix~\ref{app:failure_catalog}: Failure Catalog by Data Type}
    \item \textbf{Appendix~\ref{app:intervention}: Code Decomposition Interventions}
\end{itemize}

%===============================================================================
\clearpage
\onecolumn
\section{HumanEval and CruxEval: Prompt Examples}
\label{app:humaneval_cruxeval_prompts}
%===============================================================================

We include representative CWM prompt templates used for trace-based evaluation on HumanEval and CruxEval.

\subsection{HumanEval (output prediction with trace)}
\Needspace{8\baselineskip}
\begin{tcblisting}{
    enhanced,
    breakable,
    colback=modernlightgray,
    colframe=navyblue,
    colbacktitle=navyblue,
    coltitle=white,
    boxrule=0.4pt,
    arc=1pt,
    left=0.8mm,
    right=0.8mm,
    top=0.6mm,
    bottom=0.6mm,
    listing only,
    listing engine=listings,
    title={CWM Prompt (HumanEval)},
    fonttitle=\bfseries,
    listing options={
        basicstyle=\ttfamily\tiny,
        breaklines=true,
        breakatwhitespace=false,
        columns=fullflexible,
        frame=none,
        showstringspaces=false
    }
}
Given a python code function and an assert statement containing a specific input, provide the assertion with the exact literal output that the function returns with that input. Do not include any mathematical expressions or function calls -- only the final literal value. Your response should be solely the assertion, enclosed within [ANSWER] and [/ANSWER] tags.

You are a computational world model and can predict the program execution.
Your execution trace prediction format MUST follow this structure:
1. The execution trace prediction starts with the <|trace_context_start|> token and ends with a final <|frame_sep|> token.
2. For each code execution step:
   - Begin with <|frame_sep|> followed by the event token which can be <|call_sep|>, <|line_sep|>, <|return_sep|> or <|exception_sep|>.
   - After <|call_sep|> or <|line_sep|> put the local variable states as dictionary in JSON format followed by the <|action_sep|> token and the current source code line.
   - After <|return_sep|>, <|exception_sep|> directly put the <|action_sep|> token and the current source code line followed by an <|arg_sep|> token and the return or exception arguments.
3. Provide the full assertion with the correct output that you obtained after <|return_sep|> in [ANSWER] and [/ANSWER] tags

Here is an example of how you would predict the output of the program using your trace prediction capability:

Python function:
def f(a,b):
    y = a
    for i in range(b):
        y += y * i
    return y
assert f(1,3) == ??

Let's verify this by putting the code into a trace context and call the function in the main() function and then trace the execution of the main function.
We indicate the entry point of the execution trace with a # << START_OF_TRACE marker.

def f(a,b):
    y = a
    for i in range(b):
        y += y * i
    return y

def main(): # << START_OF_TRACE
    return f(1,3)

<|frame_sep|><|call_sep|>{}<|action_sep|>def main(): # << START_OF_TRACE
<|frame_sep|><|line_sep|>{}<|action_sep|>    return f(1,3)
<|frame_sep|><|call_sep|>{"a": "1", "b": "3"}<|action_sep|>def f(a,b):
<|frame_sep|><|line_sep|>{"a": "..", "b": ".."}<|action_sep|>    y = a
<|frame_sep|><|line_sep|>{"a": "..", "b": "..", "y": "1"}<|action_sep|>    for i in range(b):
<|frame_sep|><|line_sep|>{"a": "..", "b": "..", "y": "..", "i": "0"}<|action_sep|>        y += y * i
<|frame_sep|><|line_sep|>{"a": "..", "b": "..", "y": "..", "i": ".."}<|action_sep|>    for i in range(b):
<|frame_sep|><|line_sep|>{"a": "..", "b": "..", "y": "..", "i": "1"}<|action_sep|>        y += y * i
<|frame_sep|><|line_sep|>{"a": "..", "b": "..", "y": "2", "i": ".."}<|action_sep|>    for i in range(b):
<|frame_sep|><|line_sep|>{"a": "..", "b": "..", "y": "..", "i": "2"}<|action_sep|>        y += y * i
<|frame_sep|><|line_sep|>{"a": "..", "b": "..", "y": "6", "i": ".."}<|action_sep|>    for i in range(b):
<|frame_sep|><|line_sep|>{"a": "..", "b": "..", "y": "..", "i": ".."}<|action_sep|>    return y
<|frame_sep|><|return_sep|><|action_sep|>    return y
<|arg_sep|>"6"<|frame_sep|><|return_sep|><|action_sep|>    return f(1,3)
<|arg_sep|>"6"<|frame_sep|>

Now let us analyze the trace. The return argument of the function call f(1,3) in the main() function is "6" in JSON format, so the return value is 6.

[ANSWER]
assert f(1,3) == 6
[/ANSWER]

Now solve this problem:

Python function:
<FUNCTION_CODE>
assert <FUNCTION_NAME>(<INPUT_ARGS>) == ??

Let's verify this by putting the code into a trace context and call the function in the main() function and then trace the execution of the main function.
We indicate the entry point of the execution trace with a # << START_OF_TRACE marker.

<FUNCTION_CODE>

def main(): # << START_OF_TRACE
    return <FUNCTION_NAME>(<INPUT_ARGS>)
\end{tcblisting}

\subsection{CruxEval (output prediction with trace)}
\Needspace{8\baselineskip}
\begin{tcblisting}{
    enhanced,
    breakable,
    colback=modernlightgray,
    colframe=navyblue,
    colbacktitle=navyblue,
    coltitle=white,
    boxrule=0.4pt,
    arc=1pt,
    left=0.8mm,
    right=0.8mm,
    top=0.6mm,
    bottom=0.6mm,
    listing only,
    listing engine=listings,
    title={CWM Prompt (CruxEval-O)},
    fonttitle=\bfseries,
    listing options={
        basicstyle=\ttfamily\tiny,
        breaklines=true,
        breakatwhitespace=false,
        columns=fullflexible,
        frame=none,
        showstringspaces=false
    }
}
Given a python code function and an assert statement containing a specific input, provide the assertion with the exact literal output that the function returns with that input. Do not include any mathematical expressions or function calls -- only the final literal value. Your response should be solely the assertion, enclosed within [ANSWER] and [/ANSWER] tags.

You are a computational world model and can predict the program execution.
Your execution trace prediction format MUST follow this structure:
1. The execution trace prediction starts with the <|trace_context_start|> token and ends with a final <|frame_sep|> token.
2. For each code execution step:
   - Begin with <|frame_sep|> followed by the event token which can be <|call_sep|>, <|line_sep|>, <|return_sep|> or <|exception_sep|>.
   - After <|call_sep|> or <|line_sep|> put the local variable states as dictionary in JSON format followed by the <|action_sep|> token and the current source code line.
   - After <|return_sep|>, <|exception_sep|> directly put the <|action_sep|> token and the current source code line followed by an <|arg_sep|> token and the return or exception arguments.
3. Provide the full assertion with the correct output that you obtained after <|return_sep|> in [ANSWER] and [/ANSWER] tags

Now solve this problem:

Python function:
<FUNCTION_CODE>
assert f(<INPUT>) == ??

Let's verify this by putting the code into a trace context and call the function in the main() function and then trace the execution of the main function.
We indicate the entry point of the execution trace with a # << START_OF_TRACE marker.

<FUNCTION_CODE>

def main(): # << START_OF_TRACE
    return f(<INPUT>)
\end{tcblisting}

%===============================================================================
\section{Compositionality: Additional Details}
\label{app:atomic_accuracy}
%===============================================================================

This appendix supplements the compositionality analysis in~\cref{sec:compositionality}.

\subsection{Prompt (depth 3)}
\label{app:comp_prompt}
For reproducibility, we include an example prompt used to evaluate compositionality at depth 3.

% Keep the paragraph and the start of the box together; allow the box to span pages.
\Needspace{8\baselineskip}
\begin{tcblisting}{
    enhanced,
    breakable,
    colback=modernlightgray,
    colframe=navyblue,
    colbacktitle=navyblue,
    coltitle=white,
    boxrule=0.4pt,
    arc=1pt,
    left=0.8mm,
    right=0.8mm,
    top=0.6mm,
    bottom=0.6mm,
    listing only,
    listing engine=listings,
    title={CWM Prompt (Composition)},
    fonttitle=\bfseries,
    listing options={
        basicstyle=\ttfamily\tiny,
        breaklines=true,
        breakatwhitespace=false,
        columns=fullflexible,
        frame=none,
        showstringspaces=false
    }
}
Given a python code function and an assert statement containing a specific input, provide the assertion with the exact literal output that the function returns with that input. Do not include any mathematical expressions or function calls -- only the final literal value. Your response should be solely the assertion, enclosed within [ANSWER] and [/ANSWER] tags.

You are a computational world model and can predict the program execution.
Your execution trace prediction format MUST follow this structure:
1. The execution trace prediction starts with the <|trace_context_start|> token and ends with a final <|frame_sep|> token.
2. For each code execution step:
   - Begin with <|frame_sep|> followed by the event token which can be <|call_sep|>, <|line_sep|>, <|return_sep|> or <|exception_sep|>.
   - After <|call_sep|> or <|line_sep|> put the local variable states as dictionary in JSON format followed by the <|action_sep|> token and the current source code line.
   - After <|return_sep|>, <|exception_sep|> directly put the <|action_sep|> token and the current source code line followed by an <|arg_sep|> token and the return or exception arguments.
3. Provide the full assertion with the correct output that you obtained after <|return_sep|> in [ANSWER] and [/ANSWER] tags

Here is an example of how you would predict the output of the program using your trace prediction capability:

Python function:
def f(a,b):
    y = a
    for i in range(b):
        y += y * i
    return y
assert f(1,3) == ??

Let's verify this by putting the code into a trace context and call the function in the main() function and then trace the execution of the main function.
We indicate the entry point of the execution trace with a # << START_OF_TRACE marker.

def f(a,b):
    y = a
    for i in range(b):
        y += y * i
    return y

def main(): # << START_OF_TRACE
    return f(1,3)

<|frame_sep|><|call_sep|>{}<|action_sep|>def main(): # << START_OF_TRACE
<|frame_sep|><|line_sep|>{}<|action_sep|>    return f(1,3)
<|frame_sep|><|call_sep|>{"a": "1", "b": "3"}<|action_sep|>def f(a,b):
<|frame_sep|><|line_sep|>{"a": "..", "b": ".."}<|action_sep|>    y = a
<|frame_sep|><|line_sep|>{"a": "..", "b": "..", "y": "1"}<|action_sep|>    for i in range(b):
<|frame_sep|><|line_sep|>{"a": "..", "b": "..", "y": "..", "i": "0"}<|action_sep|>        y += y * i
<|frame_sep|><|line_sep|>{"a": "..", "b": "..", "y": "..", "i": ".."}<|action_sep|>    for i in range(b):
<|frame_sep|><|line_sep|>{"a": "..", "b": "..", "y": "..", "i": "1"}<|action_sep|>        y += y * i
<|frame_sep|><|line_sep|>{"a": "..", "b": "..", "y": "2", "i": ".."}<|action_sep|>    for i in range(b):
<|frame_sep|><|line_sep|>{"a": "..", "b": "..", "y": "..", "i": "2"}<|action_sep|>        y += y * i
<|frame_sep|><|line_sep|>{"a": "..", "b": "..", "y": "6", "i": ".."}<|action_sep|>    for i in range(b):
<|frame_sep|><|line_sep|>{"a": "..", "b": "..", "y": "..", "i": ".."}<|action_sep|>    return y
<|frame_sep|><|return_sep|><|action_sep|>    return y
<|arg_sep|>"6"<|frame_sep|><|return_sep|><|action_sep|>    return f(1,3)
<|arg_sep|>"6"<|frame_sep|>

Now let us analyze the trace. The return argument of the function call f(1,3) in the main() function is "6" in JSON format, so the return value is 6.

[ANSWER]
assert f(1,3) == 6
[/ANSWER]

Now solve this problem:

Python functions:
def func_1(s, pre):
    return pre + s

def func_12(s):
    return ''.join(ch.lower() if i % 2 == 0 else ch.upper() for i, ch in enumerate(s))

def func_14(s, sep):
    return sep.join(s)


assert main_solution("qgjucy") == ??

Let's verify this by putting the code into a trace context and call the function in the main() function and then trace the execution of the main function.
We indicate the entry point of the execution trace with a # << START_OF_TRACE marker.

def func_1(s, pre):
    return pre + s

def func_12(s):
    return ''.join(ch.lower() if i % 2 == 0 else ch.upper() for i, ch in enumerate(s))

def func_14(s, sep):
    return sep.join(s)


def main_solution(x):
    return func_12(func_12(func_14(x, '-')))

def main(): # << START_OF_TRACE
    return main_solution("qgjucy")
\end{tcblisting}

\subsection{Atomic Sring function accuracy distribution}
\label{app:atomic_string_functions}
\begin{figure}[t]
\centering
\includegraphics[width=\linewidth]{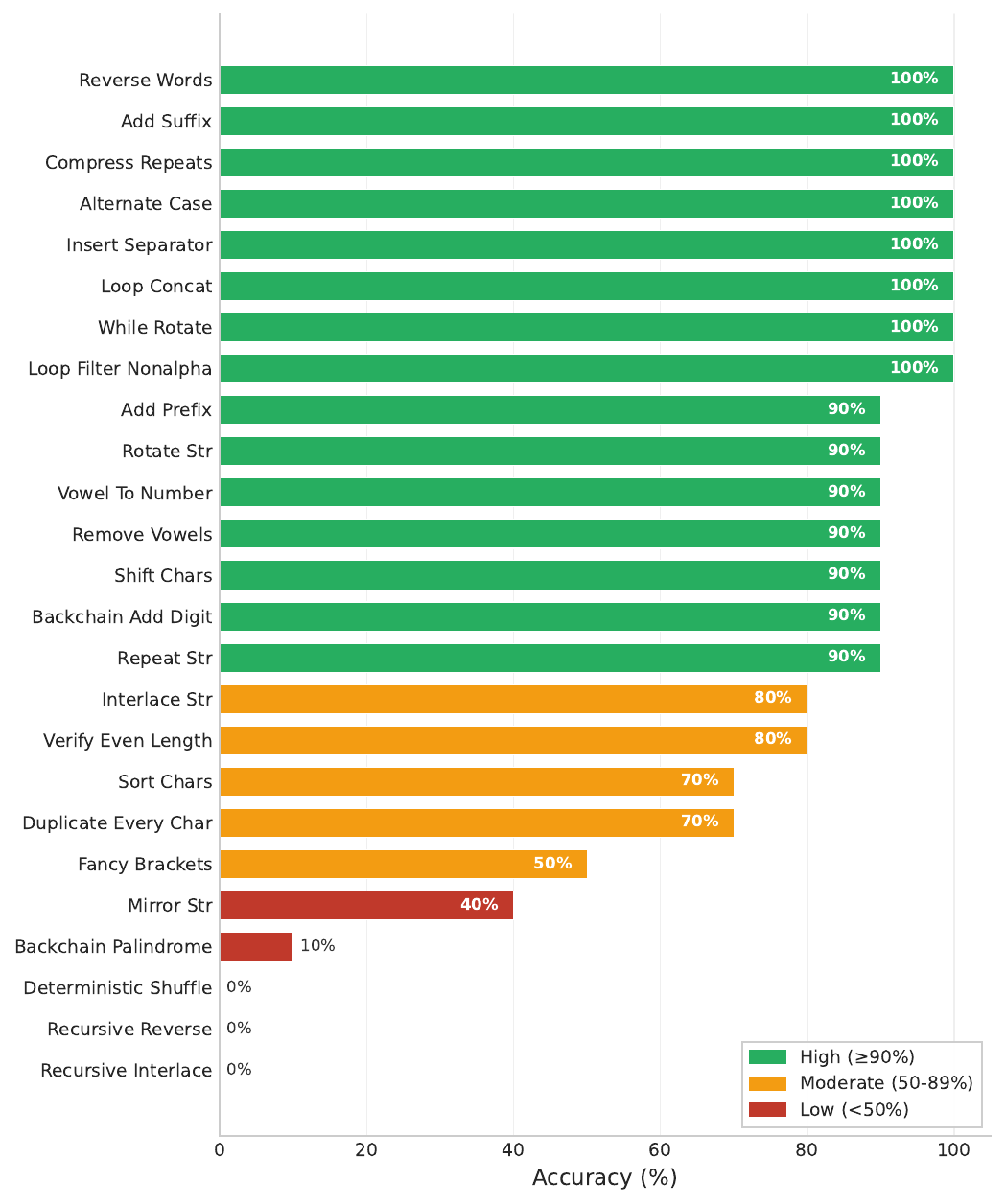}
\caption{Atomic accuracy report for the 25 string-manipulation functions used in the compositionality study.}
\label{fig:atomic_all25_accuracy}
\end{figure}

Before measuring composition depth effects, we evaluate all 25 atomic string functions to separate intrinsic operation difficulty from compositional degradation.
\Cref{fig:atomic_all25_accuracy} summarizes the per-function atomic accuracies used to select the high-accuracy subset for the depth experiment.

\subsection{Composition Zoo function definitions by category}
\label{app:composition_zoo_function_defs}
For the multi-domain compositionality study (``Composition Zoo''), each category uses five deterministic, single-argument functions that are composed to depth 5.
We list the exact function definitions used to generate the evaluation prompts.

\Needspace{6\baselineskip}
\begin{tcblisting}{
    enhanced,
    breakable,
    colback=modernlightgray,
    colframe=navyblue,
    colbacktitle=navyblue,
    coltitle=white,
    boxrule=0.4pt,
    arc=1pt,
    left=0.8mm,
    right=0.8mm,
    top=0.6mm,
    bottom=0.6mm,
    listing only,
    listing engine=listings,
    title={Composition Zoo Functions (Boolean)},
    fonttitle=\bfseries,
    listing options={
        language=Python,
        basicstyle=\ttfamily\scriptsize,
        breaklines=true,
        breakatwhitespace=false,
        columns=fullflexible,
        frame=none,
        showstringspaces=false
    }
}
def bool_and_true(x):
    return x and True

def bool_or_false(x):
    return x or False

def bool_not(x):
    return not x

def bool_identity(x):
    return x

def bool_xor_true(x):
    return x != True
\end{tcblisting}

\Needspace{6\baselineskip}
\begin{tcblisting}{
    enhanced,
    breakable,
    colback=modernlightgray,
    colframe=navyblue,
    colbacktitle=navyblue,
    coltitle=white,
    boxrule=0.4pt,
    arc=1pt,
    left=0.8mm,
    right=0.8mm,
    top=0.6mm,
    bottom=0.6mm,
    listing only,
    listing engine=listings,
    title={Composition Zoo Functions (Bitwise)},
    fonttitle=\bfseries,
    listing options={
        language=Python,
        basicstyle=\ttfamily\scriptsize,
        breaklines=true,
        breakatwhitespace=false,
        columns=fullflexible,
        frame=none,
        showstringspaces=false
    }
}
def bit_and_15(x):
    return x & 15

def bit_or_8(x):
    return x | 8

def bit_xor_7(x):
    return x ^ 7

def bit_shift_left_1(x):
    return x << 1

def bit_shift_right_1(x):
    return x >> 1
\end{tcblisting}

\Needspace{6\baselineskip}
\begin{tcblisting}{
    enhanced,
    breakable,
    colback=modernlightgray,
    colframe=navyblue,
    colbacktitle=navyblue,
    coltitle=white,
    boxrule=0.4pt,
    arc=1pt,
    left=0.8mm,
    right=0.8mm,
    top=0.6mm,
    bottom=0.6mm,
    listing only,
    listing engine=listings,
    title={Composition Zoo Functions (Math)},
    fonttitle=\bfseries,
    listing options={
        language=Python,
        basicstyle=\ttfamily\scriptsize,
        breaklines=true,
        breakatwhitespace=false,
        columns=fullflexible,
        frame=none,
        showstringspaces=false
    }
}
def math_abs(x):
    return abs(x)

def math_negate(x):
    return -x

def math_double(x):
    return x * 2

def math_halve(x):
    return x // 2

def math_mod_10(x):
    return x % 10
\end{tcblisting}

\Needspace{6\baselineskip}
\begin{tcblisting}{
    enhanced,
    breakable,
    colback=modernlightgray,
    colframe=navyblue,
    colbacktitle=navyblue,
    coltitle=white,
    boxrule=0.4pt,
    arc=1pt,
    left=0.8mm,
    right=0.8mm,
    top=0.6mm,
    bottom=0.6mm,
    listing only,
    listing engine=listings,
    title={Composition Zoo Functions (Character)},
    fonttitle=\bfseries,
    listing options={
        language=Python,
        basicstyle=\ttfamily\scriptsize,
        breaklines=true,
        breakatwhitespace=false,
        columns=fullflexible,
        frame=none,
        showstringspaces=false
    }
}
def char_next(c):
    return chr((ord(c) - ord('a') + 1) % 26 + ord('a'))

def char_prev(c):
    return chr((ord(c) - ord('a') - 1) % 26 + ord('a'))

def char_shift_3(c):
    return chr((ord(c) - ord('a') + 3) % 26 + ord('a'))

def char_shift_5(c):
    return chr((ord(c) - ord('a') + 5) % 26 + ord('a'))

def char_identity(c):
    return c
\end{tcblisting}

\Needspace{6\baselineskip}
\begin{tcblisting}{
    enhanced,
    breakable,
    colback=modernlightgray,
    colframe=navyblue,
    colbacktitle=navyblue,
    coltitle=white,
    boxrule=0.4pt,
    arc=1pt,
    left=0.8mm,
    right=0.8mm,
    top=0.6mm,
    bottom=0.6mm,
    listing only,
    listing engine=listings,
    title={Composition Zoo Functions (List)},
    fonttitle=\bfseries,
    listing options={
        language=Python,
        basicstyle=\ttfamily\scriptsize,
        breaklines=true,
        breakatwhitespace=false,
        columns=fullflexible,
        frame=none,
        showstringspaces=false
    }
}
def list_append_0(lst):
    return lst + [0]

def list_prepend_1(lst):
    return [1] + lst

def list_reverse(lst):
    return lst[::-1]

def list_drop_first(lst):
    return lst[1:] if len(lst) > 1 else lst

def list_drop_last(lst):
    return lst[:-1] if len(lst) > 1 else lst
\end{tcblisting}

\Needspace{6\baselineskip}
\begin{tcblisting}{
    enhanced,
    breakable,
    colback=modernlightgray,
    colframe=navyblue,
    colbacktitle=navyblue,
    coltitle=white,
    boxrule=0.4pt,
    arc=1pt,
    left=0.8mm,
    right=0.8mm,
    top=0.6mm,
    bottom=0.6mm,
    listing only,
    listing engine=listings,
    title={Composition Zoo Functions (Set)},
    fonttitle=\bfseries,
    listing options={
        language=Python,
        basicstyle=\ttfamily\scriptsize,
        breaklines=true,
        breakatwhitespace=false,
        columns=fullflexible,
        frame=none,
        showstringspaces=false
    }
}
def set_add_1(s):
    return s | {1}

def set_add_2(s):
    return s | {2}

def set_remove_1(s):
    return s - {1}

def set_remove_2(s):
    return s - {2}

def set_intersect_123(s):
    return s & {1, 2, 3}
\end{tcblisting}

\Needspace{6\baselineskip}
\begin{tcblisting}{
    enhanced,
    breakable,
    colback=modernlightgray,
    colframe=navyblue,
    colbacktitle=navyblue,
    coltitle=white,
    boxrule=0.4pt,
    arc=1pt,
    left=0.8mm,
    right=0.8mm,
    top=0.6mm,
    bottom=0.6mm,
    listing only,
    listing engine=listings,
    title={Composition Zoo Functions (Dictionary)},
    fonttitle=\bfseries,
    listing options={
        language=Python,
        basicstyle=\ttfamily\scriptsize,
        breaklines=true,
        breakatwhitespace=false,
        columns=fullflexible,
        frame=none,
        showstringspaces=false
    }
}
def dict_set_a_1(d):
    return {**d, 'a': 1}

def dict_set_b_2(d):
    return {**d, 'b': 2}

def dict_remove_a(d):
    return {k: v for k, v in d.items() if k != 'a'}

def dict_remove_b(d):
    return {k: v for k, v in d.items() if k != 'b'}

def dict_inc_a(d):
    return {**d, 'a': d.get('a', 0) + 1}
\end{tcblisting}

\Needspace{10\baselineskip}
\begin{tcblisting}{
    enhanced,
    breakable,
    colback=modernlightgray,
    colframe=navyblue,
    colbacktitle=navyblue,
    coltitle=white,
    boxrule=0.4pt,
    arc=1pt,
    left=0.8mm,
    right=0.8mm,
    top=0.6mm,
    bottom=0.6mm,
    listing only,
    listing engine=listings,
    title={Composition Zoo Functions (String)},
    fonttitle=\bfseries,
    listing options={
        language=Python,
        basicstyle=\ttfamily\tiny,
        breaklines=true,
        breakatwhitespace=false,
        columns=fullflexible,
        frame=none,
        showstringspaces=false
    }
}
def reverse_words(s):
    words = s.split()
    return ' '.join(reversed(words))

def add_suffix(s, suf):
    return s + suf

def compress_repeats(s):
    if not s:
        return s
    result = [s[0]]
    for ch in s[1:]:
        if ch != result[-1]:
            result.append(ch)
    return ''.join(result)

def alternate_case(s):
    return ''.join(ch.lower() if i % 2 == 0 else ch.upper() for i, ch in enumerate(s))

def insert_separator(s, sep):
    return sep.join(s)

def loop_concat(s, n):
    result = ""
    for _ in range(n):
        result += s
    return result

def while_rotate(s, n):
    count = 0
    while count < n and s:
        s = s[1:] + s[0]
        count += 1
    return s

def loop_filter_nonalpha(s):
    result = ""
    for ch in s:
        if ch.isalpha():
            result += ch
    return result

def add_prefix(s, pre):
    return pre + s

def rotate_str(s, n):
    if not s:
        return s
    n = n % len(s)
    return s[n:] + s[:n]

def vowel_to_number(s):
    mapping = {
        'a': '1', 'e': '2', 'i': '3', 'o': '4', 'u': '5',
        'A': '1', 'E': '2', 'I': '3', 'O': '4', 'U': '5'
    }
    return ''.join(mapping.get(ch, ch) for ch in s)

def remove_vowels(s):
    vowels = 'aeiouAEIOU'
    return ''.join(ch for ch in s if ch not in vowels)

def shift_chars(s, shift):
    def shift_char(ch):
        if 'a' <= ch <= 'z':
            return chr((ord(ch) - ord('a') + shift) % 26 + ord('a'))
        elif 'A' <= ch <= 'Z':
            return chr((ord(ch) - ord('A') + shift) % 26 + ord('A'))
        return ch

    return ''.join(shift_char(ch) for ch in s)

def backchain_add_digit(s, depth):
    def has_digit(t):
        return any(ch.isdigit() for ch in t)

    transformations = [
        lambda t: t + "1",
        lambda t: "2" + t,
        lambda t: t.replace("a", "3"),
        lambda t: t[::-1],
    ]

    def helper(t, d):
        if has_digit(t):
            return t
        if d == 0:
            return None
        for trans in transformations:
            new_t = trans(t)
            res = helper(new_t, d - 1)
            if res is not None:
                return res
        return None

    result = helper(s, depth)
    return result if result is not None else s

def repeat_str(s, n):
    return s * n
\end{tcblisting}

%===============================================================================
\clearpage
\section{Code $S_5$ Long-Horizon State-Tracking: Experiment Setup and Prompts (GPT-5 and CWM)}
\label{app:llm_evaluation_prompts}
%===============================================================================

This appendix documents the evaluation setup and the prompts used for the controlled $S_5$ REPL-trace benchmark (\cref{fig:main_results} and \cref{fig:teacher_forcing_process}).

\textbf{Task.}
Each example initializes five variables $(a,b,c,d,e)$ with integer values and applies $N$ swap operations implemented as Python simultaneous assignment.
We evaluate lengths $N\in\{8,16,32,64,128\}$.
Accuracy is exact match of the final variable assignment.

\textbf{Models and interfaces.}
GPT-5 is evaluated via a structured chat prompt and is required to emit only the final assignment.
CWM is evaluated in its native trace format and generates an execution trace with explicit JSON states at each step; we extract the last state frame for scoring.

\subsection{GPT-5 prompt}
\label{app:gpt5_prompt}

\Needspace{8\baselineskip}
\begin{tcblisting}{
    enhanced,
    breakable,
    colback=modernlightgray,
    colframe=navyblue,
    colbacktitle=navyblue,
    coltitle=white,
    boxrule=0.4pt,
    arc=1pt,
    left=0.8mm,
    right=0.8mm,
    top=0.6mm,
    bottom=0.6mm,
    listing only,
    listing engine=listings,
    title={GPT-5 Prompt (S5)},
    fonttitle=\bfseries,
    listing options={
        basicstyle=\ttfamily\tiny,
        breaklines=true,
        breakatwhitespace=false,
        columns=fullflexible,
        frame=none,
        showstringspaces=false
    }
}
### System prompt
You are a Python code execution tracer. Your task is to trace through Python code that performs variable assignments and swaps, then determine the final values of ALL variables.

## Task Description
Given a Python function that:
1. Initializes 5 variables (a, b, c, d, e) with integer values
2. Performs a series of simultaneous variable swaps (e.g., a, b, c, d, e = c, e, b, a, d)

You must trace through all the operations step by step and provide the final values of ALL five variables.

## Example
Code:
def execute_repl_trace():
    a = 1
    b = 2
    c = 3
    d = 4
    e = 5
    a, b, c, d, e = c, e, b, a, d
    a, b, c, d, e = e, b, c, d, a

def main():
    execute_repl_trace()

Step-by-step trace:
1. Initial: a=1, b=2, c=3, d=4, e=5
2. After a, b, c, d, e = c, e, b, a, d: a=3, b=5, c=2, d=1, e=4
3. After a, b, c, d, e = e, b, c, d, a: a=4, b=5, c=2, d=1, e=3

Answer: a=4,b=5,c=2,d=1,e=3

## Instructions
- Trace through each assignment carefully
- Remember that tuple unpacking in Python happens simultaneously (all right-hand values are evaluated before any assignment)
- Provide the final values of ALL variables in the format: a=X,b=X,c=X,d=X,e=X
- Do not include any explanation, just the comma-separated values

### User prompt (example with 8 swap operations)
Trace through the following Python code and provide the final values of ALL variables.

def execute_repl_trace():
    """Execute the REPL trace operations."""
    a = 8
    b = 4
    c = 7
    d = 8
    e = 7
    a, b, c, d, e = c, e, b, a, d
    a, b, c, d, e = e, b, c, d, a
    a, b, c, d, e = b, e, a, c, d
    a, b, c, d, e = a, b, e, d, c
    a, b, c, d, e = b, c, e, a, d
    a, b, c, d, e = e, a, c, b, d
    a, b, c, d, e = a, e, c, b, d
    a, b, c, d, e = b, d, e, c, a

def main():
    execute_repl_trace()

What are the final values of all variables? Provide in the format: a=X,b=X,c=X,d=X,e=X

\end{tcblisting}

\subsection{CWM prompt (native trace format)}
\label{app:cwm_prompt}

\Needspace{8\baselineskip}
\begin{tcblisting}{
    enhanced,
    breakable,
    colback=modernlightgray,
    colframe=navyblue,
    colbacktitle=navyblue,
    coltitle=white,
    boxrule=0.4pt,
    arc=1pt,
    left=0.8mm,
    right=0.8mm,
    top=0.6mm,
    bottom=0.6mm,
    listing only,
    listing engine=listings,
    title={CWM Prompt (S5)},
    fonttitle=\bfseries,
    listing options={
        basicstyle=\ttfamily\tiny,
        breaklines=true,
        breakatwhitespace=false,
        columns=fullflexible,
        frame=none,
        showstringspaces=false
    }
}
<|begin_of_text|><|trace_context_start|>
def execute_repl_trace():
    """Execute the REPL trace operations."""
    a = 8
    b = 4
    c = 7
    d = 8
    e = 7
    a, b, c, d, e = c, e, b, a, d
    a, b, c, d, e = e, b, c, d, a
    a, b, c, d, e = b, e, a, c, d
    a, b, c, d, e = a, b, e, d, c
    a, b, c, d, e = b, c, e, a, d
    a, b, c, d, e = e, a, c, b, d
    a, b, c, d, e = a, e, c, b, d
    a, b, c, d, e = b, d, e, c, a
    print(f"c = {c}")

def main(): # << START_OF_TRACE
    execute_repl_trace()
<|frame_sep|>
\end{tcblisting}

%===============================================================================
\clearpage
\section{Data Type Distribution Analysis}
\label{app:datatype_distribution}
%===============================================================================

Here we provide the full data type distribution for both CruxEval-Output and HumanEval, comparing the distribution in the benchmark to the distribution among wrong answer failures (excluding truncation cases).

\begin{table}[h]
\centering
\caption{CruxEval-O: Data type distribution in benchmark vs. wrong answer failures (n=800 samples, 114 wrong answer failures)}
\label{tab:datatype_cruxeval}
\begin{tabular}{lccc}
\toprule
\textbf{Type} & \textbf{Benchmark \%} & \textbf{Failure \%} & \textbf{Ratio} \\
\midrule
str   & 46.4\% & 72.8\% & 1.57$\times$ \\
list  & 24.6\% & 8.8\%  & 0.36$\times$ \\
int   & 12.1\% & 11.4\% & 0.94$\times$ \\
dict  & 8.4\%  & 0.0\%  & ---          \\
bool  & 6.1\%  & 3.5\%  & 0.57$\times$ \\
tuple & 2.0\%  & 2.6\%  & 1.32$\times$ \\
bytes & 0.2\%  & 0.9\%  & 3.51$\times$ \\
float & 0.1\%  & 0.0\%  & ---          \\
\bottomrule
\end{tabular}
\end{table}

\begin{table}[h]
\centering
\caption{HumanEval: Data type distribution in benchmark vs. wrong answer failures (n=723 samples, 32 wrong answer failures)}
\label{tab:datatype_humaneval}
\begin{tabular}{lccc}
\toprule
\textbf{Type} & \textbf{Benchmark \%} & \textbf{Failure \%} & \textbf{Ratio} \\
\midrule
str      & 17.2\% & 44.0\% & 2.56$\times$ \\
list     & 24.6\% & 25.0\% & 1.02$\times$ \\
int      & 31.5\% & 18.8\% & 0.60$\times$ \\
bool     & 19.2\% & 6.2\%  & 0.32$\times$ \\
float    & 1.4\%  & 6.2\%  & 4.52$\times$ \\
tuple    & 4.7\%  & 0.0\%  & ---          \\
dict     & 0.7\%  & 0.0\%  & ---          \\
NoneType & 0.7\%  & 0.0\%  & ---          \\
\bottomrule
\end{tabular}
\end{table}

The ``Ratio'' column shows the overrepresentation factor: values $>1$ indicate the data type appears more frequently in wrong answer failures than expected from its benchmark frequency.
In CruxEval, strings are 1.57$\times$ overrepresented in wrong answer failures (72.8\% of failures vs 46.4\% of benchmark).
In HumanEval, string-related failures are 2.56$\times$ overrepresented (44\% of failures vs 17.2\% of benchmark).

%===============================================================================
\clearpage
\section{Failure Catalog by Data Type}
\label{app:failure_catalog}
%===============================================================================

This appendix provides a detailed catalog of CruxEval-O failure cases, organized by output data type.
Each entry shows the function code, input, expected output, and CWM's incorrect prediction.

%-------------------------------------------------------------------------------
\subsection{String Failures (71\% of CruxEval-O failures)}
%-------------------------------------------------------------------------------

String operations dominate failures due to method semantics errors and tokenization brittleness.

\subsubsection{Method Semantics Errors}

\begin{table}[h]
\centering
\footnotesize
\setlength{\tabcolsep}{3pt}
\begin{tabular}{@{}p{0.12\linewidth}p{0.38\linewidth}p{0.22\linewidth}p{0.22\linewidth}@{}}
\toprule
\textbf{Sample} & \textbf{Code Snippet} & \textbf{Expected} & \textbf{Predicted} \\
\midrule
\texttt{s\_113} \newline \texttt{swapcase} &
\texttt{if count\%2==0:} \newline \texttt{~~a.append(line[i].swapcase())} &
\texttt{'987YhnShAShD...'} &
\texttt{'987YhNShAshD...'} \\
\midrule
\texttt{s\_114} \newline \texttt{rsplit} &
\texttt{text.rsplit(sep, maxsplit=2)} &
\texttt{['a', '', 'b']} &
\texttt{['a', '-', 'b']} \\
\midrule
\texttt{s\_136} \newline \texttt{center} &
\texttt{line.center(width)} &
\texttt{'~~bc~'} &
\texttt{'~bc~~'} \\
\midrule
\texttt{s\_23} \newline \texttt{rstrip} &
\texttt{text.rstrip(chars)} &
\texttt{'...XQuery 2.'} &
\texttt{'...XQuery 2.2'} \\
\midrule
\texttt{s\_239} \newline \texttt{lstrip/rstrip} &
\texttt{text.lstrip(froms)} \newline \texttt{text.rstrip(froms)} &
\texttt{'1co'} &
\texttt{'t 1cos'} \\
\midrule
\texttt{s\_168} \newline \texttt{translate} &
\texttt{text.translate(key)} &
\texttt{'spaib'} &
\texttt{'spai'} \\
\midrule
\texttt{s\_211} \newline \texttt{rindex/index} &
\texttt{if s.rindex(c) != s.index(c):} \newline \texttt{~~count+=1} &
\texttt{10} &
\texttt{11} \\
\bottomrule
\end{tabular}
\caption{String method semantics errors. Model misunderstands Python built-in method behavior.}
\label{tab:string_method_errors}
\end{table}

\subsubsection{Index and Slice Errors}

\begin{table}[h]
\centering
\footnotesize
\setlength{\tabcolsep}{3pt}
\begin{tabular}{@{}p{0.12\linewidth}p{0.38\linewidth}p{0.22\linewidth}p{0.22\linewidth}@{}}
\toprule
\textbf{Sample} & \textbf{Code Snippet} & \textbf{Expected} & \textbf{Predicted} \\
\midrule
\texttt{s\_198} \newline \texttt{reverse+strip} &
\texttt{text[::-1].strip(chars)[::-1]} &
\texttt{'tcmfsm'} &
\texttt{'tcmfs'} \\
\midrule
\texttt{s\_201} \newline \texttt{reverse digits} &
\texttt{chars[::-1]} after \texttt{isdigit()} filter &
\texttt{'641524'} &
\texttt{'61452'} \\
\midrule
\texttt{s\_218} \newline \texttt{count+reverse} &
\texttt{((string+sep) * cnt)[::-1]} &
\texttt{'bacfbacfcbaac...'} &
\texttt{'cfbafcabcfcaab...'} \\
\midrule
\texttt{s\_220} \newline \texttt{format+slice} &
\texttt{text[:m]}, \texttt{text[n:]} in format &
\texttt{'bagfedcacbagfedc'} &
\texttt{'cbagfeddaacbafedc'} \\
\midrule
\texttt{s\_237} \newline \texttt{partition} &
\texttt{text.partition(char)} with slice &
\texttt{'uuzlwaqiaj'} &
\texttt{'zlwaqiaj'} \\
\bottomrule
\end{tabular}
\caption{String indexing and slicing errors involving reverse operations and complex slice expressions.}
\label{tab:string_slice_errors}
\end{table}

%-------------------------------------------------------------------------------
\subsection{List Failures (12\% of CruxEval-O failures)}
%-------------------------------------------------------------------------------

List failures primarily involve mutation during iteration.

\begin{table}[h]
\centering
\footnotesize
\setlength{\tabcolsep}{3pt}
\begin{tabular}{@{}p{0.12\linewidth}p{0.42\linewidth}p{0.18\linewidth}p{0.22\linewidth}@{}}
\toprule
\textbf{Sample} & \textbf{Code Snippet} & \textbf{Expected} & \textbf{Predicted} \\
\midrule
\texttt{s\_112} \newline \texttt{remove in loop} &
\texttt{for letter in ls:} \newline \texttt{~~if not letter.istitle():} \newline \texttt{~~~~ls.remove(letter)} &
\texttt{'XYZLtRRdn...'} &
\texttt{'XYZLRergH...'} \\
\midrule
\texttt{s\_149} \newline \texttt{popitem loop} &
\texttt{dict.fromkeys(...).popitem()[0]} in loop &
\texttt{'2,4,2,0,'} &
\texttt{'2,,,,,3,,,,,1...'} \\
\midrule
\texttt{s\_150} \newline \texttt{insert loop} &
\texttt{for n in numbers[index:]:} \newline \texttt{~~numbers.insert(index, n)} &
\texttt{[-2, 4, -4]} &
\texttt{null (truncated)} \\
\bottomrule
\end{tabular}
\caption{List mutation errors. Modifying a list while iterating causes tracking failures or infinite loops.}
\label{tab:list_errors}
\end{table}

%-------------------------------------------------------------------------------
\subsection{Integer Failures (8\% of CruxEval-O failures)}
%-------------------------------------------------------------------------------

Integer failures involve counter tracking and range calculation errors.

\begin{table}[h]
\centering
\footnotesize
\setlength{\tabcolsep}{3pt}
\begin{tabular}{@{}p{0.12\linewidth}p{0.42\linewidth}p{0.18\linewidth}p{0.22\linewidth}@{}}
\toprule
\textbf{Sample} & \textbf{Code Snippet} & \textbf{Expected} & \textbf{Predicted} \\
\midrule
\texttt{s\_118} \newline \texttt{range loop} &
\texttt{for i in range(num\_applies):} \newline \texttt{~~extra\_chars += chars} &
\texttt{'zbzquiuqnmfkx'} &
\texttt{'zbzquiuqnfkx'} \\
\midrule
\texttt{s\_163} \newline \texttt{dynamic range} &
\texttt{range(size-len(text))} &
\texttt{'w))))))))))))'} &
\texttt{'w)))))))'} \\
\midrule
\texttt{s\_244} \newline \texttt{rjust count} &
\texttt{text.rjust(len(text) + count*2)} &
\texttt{'~~~~~~~~'} &
\texttt{'~~~~~~'} \\
\bottomrule
\end{tabular}
\caption{Integer counter and range errors. Model loses track of iteration counts or range bounds.}
\label{tab:integer_errors}
\end{table}

%-------------------------------------------------------------------------------
\subsection{Dictionary Failures}
%-------------------------------------------------------------------------------

\begin{table}[h]
\centering
\footnotesize
\setlength{\tabcolsep}{3pt}
\begin{tabular}{@{}p{0.12\linewidth}p{0.42\linewidth}p{0.18\linewidth}p{0.22\linewidth}@{}}
\toprule
\textbf{Sample} & \textbf{Code Snippet} & \textbf{Expected} & \textbf{Predicted} \\
\midrule
\texttt{s\_130} \newline \texttt{items+format} &
\texttt{items = list(m.items())} \newline swap loop + \texttt{.format(*m.keys(), **m)} &
\texttt{'h=l'} &
\texttt{'l=4'} \\
\bottomrule
\end{tabular}
\caption{Dictionary iteration and format string errors.}
\label{tab:dict_errors}
\end{table}

%-------------------------------------------------------------------------------
\subsection{Truncation Failures (8\% of CruxEval-O failures)}
%-------------------------------------------------------------------------------

Truncation occurs when execution traces exceed the token budget.

\begin{table}[h]
\centering
\footnotesize
\setlength{\tabcolsep}{3pt}
\begin{tabular}{@{}p{0.18\linewidth}p{0.50\linewidth}p{0.26\linewidth}@{}}
\toprule
\textbf{Pattern} & \textbf{Example Code} & \textbf{Cause} \\
\midrule
\textbf{List mutation} &
\texttt{for n in numbers[index:]:} \newline \texttt{~~numbers.insert(index, n)} &
Infinite loop from growing list \\
\midrule
\textbf{Nested iteration} &
\texttt{for i in range(len(text)):} \newline \texttt{~~for j in range(i, len(text)):} &
$O(n^2)$ trace expansion \\
\midrule
\textbf{Long input} &
Input string $>$30 characters with per-char loop &
Dense trace per character \\
\bottomrule
\end{tabular}
\caption{Truncation patterns. Dense state snapshots cause traces to exceed the 8K token budget.}
\label{tab:truncation_patterns}
\end{table}

%===============================================================================
\clearpage
\section{Code Decomposition Interventions}
\label{app:intervention}
%===============================================================================

The intervention rewrites programs so that intermediate values implicit in standard Python syntax become explicit variables in the execution trace.
This allows CWMs to condition on intermediate dummy variables at each generated command.
We implement the intervention using Python’s Abstract Syntax Tree (AST) and apply it only to failing samples.

We extract complex subexpressions into temporary variables while preserving evaluation order and program semantics.
For example:
\begin{tcblisting}{
    enhanced,
    breakable,
    colback=modernlightgray,
    colframe=navyblue,
    boxrule=0.4pt,
    arc=1pt,
    left=1mm,
    right=1mm,
    top=0.6mm,
    bottom=0.6mm,
    listing only,
    listing engine=listings,
    title={Expression Decomposition},
    fonttitle=\bfseries,
    listing options={
        language=Python,
        basicstyle=\ttfamily\scriptsize,
        breaklines=true,
        showstringspaces=false,
        frame=none
    }
}
# Before:
output.append((nums.count(n), n))

# After:
_t0 = nums.count(n)
_t1 = (_t0, n)
output.append(_t1)
\end{tcblisting}

We traverse the AST and identify subexpressions whose structural complexity exceeds a threshold.
For each such subexpression, we:
\begin{enumerate}[nosep,leftmargin=*]
    \item Generate a fresh temporary variable name.
    \item Insert an assignment that extracts the subexpression.
    \item Replace the original subexpression with the temporary variable reference.
\end{enumerate}

Atomic nodes (\texttt{Name}, \texttt{Constant}) are left unchanged.
Extraction is applied to \texttt{BinOp}, \texttt{Call}, \texttt{Subscript}, \texttt{Compare}, and container literals.
Control-flow structure is preserved: extracted assignments are inserted immediately before the original statement.

\subsection{String Operation Decomposition}

String methods such as \texttt{.index}, \texttt{.find}, and \texttt{.replace} often fail when the target character is embedded inside a single BPE token, rendering the relevant character-level state invisible to the model.
To address this, we expand selected string operations into explicit character-level loops.

We decompose the following single-character string operations:
\[
\texttt{index},\ \texttt{rindex},\ \texttt{find},\ \texttt{rfind},\ \texttt{count},\ \texttt{replace},\ \texttt{c in s}
\]
Decomposition is applied only when the search target is statically verified to be a single character.

Example:
\\
\begin{tcblisting}{
    enhanced,
    breakable,
    colback=modernlightgray,
    colframe=navyblue,
    boxrule=0.4pt,
    arc=1pt,
    left=1mm,
    right=1mm,
    top=0.6mm,
    bottom=0.6mm,
    listing only,
    listing engine=listings,
    title={String Operation Decomposition (\texttt{index})},
    fonttitle=\bfseries,
    listing options={
        language=Python,
        basicstyle=\ttfamily\scriptsize,
        breaklines=true,
        showstringspaces=false,
        frame=none
    }
}
# Original:
pos = text.index(char)

# Transformed:
_t0 = list(text)
_t1 = -1
for _t2 in range(len(_t0)):
    if _t0[_t2] == char:
        _t1 = _t2
        break
if _t1 == -1:
    raise ValueError("substring not found")
pos = _t1
\end{tcblisting}

This expansion makes each character comparison explicit in the trace, restoring per-position visibility. For each rewritten program, semantic equivalence is verified by executing both the original and transformed code on the same inputs and checking for identical outputs.

\subsection{Failure Modes and Limitations}

While decomposition recovers a subset of failures, it introduces two inherent limitations.

\paragraph{Token explosion.}
Loop-based string decomposition generates one trace step per character.
For long strings, this leads to trace lengths that exceed the context window, resulting in truncation.

\paragraph{Unrecoverable errors.}
Failures caused by incorrect semantic reasoning (e.g., misunderstanding Python method behavior), long-horizon state drift through loops, or complex control flow are not corrected by decomposition, as these errors do not arise from missing intermediate visibility.

The aggregate effect of these interventions is reported in Table~\ref{tab:benchmark_results} under ``After Intervention. This result illustrates the limited but consistent improvement in controlled string composition after decomposition. We emphasize that decomposition is used here as a diagnostic tool to isolate visibility-related failures, rather than as a scalable execution strategy.

% %===============================================================================
% \clearpage
% \section{Composition Accuracy Figure}
% \label{app:composition_accuracy_figure}
% %===============================================================================

% \begin{figure}[h]
% \centering
% \includegraphics[width=0.9\linewidth]{composition_accuracy_with_flattening.pdf}
% \caption{Composition accuracy on nested string functions. Observed accuracy (blue) falls far below the theoretical baseline (dashed) computed from atomic accuracy ($0.953^d$). Flattening (red) provides modest improvement.}
% \label{fig:composition_accuracy}
% \end{figure}

\end{document}